\documentclass[showpacs,amsmath,amssymb, twocolumn,pra,superscriptaddress]{revtex4-2}
\usepackage{amsmath}
\usepackage{amsthm} %May cause conflict later
\usepackage{mathtools}
\usepackage{bm}
\usepackage{graphicx}
\usepackage{esvect}
\usepackage{graphics}
\usepackage{amssymb}
\usepackage{enumitem}
\usepackage{braket}
\usepackage{xcolor}
\usepackage[colorlinks = true]{hyperref}
\usepackage{subfigure}
\usepackage{tikz}
\usetikzlibrary{decorations.pathreplacing} %For curly braces

\usepackage[linesnumbered,ruled,vlined]{algorithm2e} %REMOVE THIS IF IT DOESNT WORK, USED FOR ALGORITHM

\definecolor{egg}{rgb}{.98,.97,.92}
\definecolor{astroorange}{rgb}{1,.93,.79}
\definecolor{darkorange}{rgb}{1,.89,.6}
\definecolor{dullblue}{rgb}{.29,.47,.77}
\definecolor{grayblue}{rgb}{.98,.98,.98}
\definecolor{fadedblue}{rgb}{.78,.86,.92}
\definecolor{tiffanyblue}{rgb}{.96,1,1}
\definecolor{grayish}{rgb}{.93,.93,.97}
\definecolor{charcoal}{rgb}{.247,.259,.27}
\definecolor{evergreen}{rgb}{.7725,.858,.7647}
\definecolor{dullred}{rgb}{.929,.498,.598}
\definecolor{lavender}{rgb}{.8,.741,.85}
\definecolor{mustard}{rgb}{1,.938,.413}
\definecolor{pastelorange}{rgb}{.995,.80,.469}

\newcommand{\tr}[1]{\mathrm{Tr}\left\{#1\right\}}

\newcommand{\av}[1]{\underset{\tiny{#1}}{\mathbb{E}}}

\newcommand{\abs}[1]{\left| #1 \right|}
\newcommand{\OTOC}{\mathrm{OTOC}}
\newcommand{\insta}{\mathbb{I}}

\newcommand{\xpos}{0} %Tikz node positions
\newcommand{\ypos}{0} %Tikz node positions
\newcommand{\xrel}{1} %Tikz node relative x distance to center
\newcommand{\yrel}{.5} %Tikz node relative y distance to center
\newcommand{\xglobalshift}{0} %Global Shift Parameter
\newcommand{\yglobalshift}{0} %Global shift parameter
\newcommand{\height}{2em} %node height
\newcommand{\width}{2em} %node width
\newcommand{\name}{} %Tikz node name
\newcommand{\nodenum}{0} %Tikz node name
\newcommand{\heightsingle}{2em} %node height
\newcommand{\heightdouble}{4.6em} %node height
\newcommand{\widthsingle}{2em} %node height
\newcommand{\rowspace}{2*\yrel} %Tikz node row space
\newtheorem{theorem}{Theorem}

\newtheorem{prop}{Proposition}
\newtheorem{corollary}{Corollary}
\newtheorem{definition}{Definition}
\newtheorem{conjecture}{Conjecture}

\begin{document}
\title{On the Hardness of Measuring Magic}
\author{Roy J. Garcia}
\email{roygarcia@g.harvard.edu}
\affiliation{Department of Physics, Harvard University, Cambridge, MA 02138, USA}

\author{Gaurav Bhole}
\email{gaurav\_bhole@fas.harvard.edu}
\affiliation{Department of Physics, Harvard University, Cambridge, MA 02138, USA}
\affiliation{Department of Cancer Biology, Dana-Farber Cancer Institute, Boston, MA, 02215 USA}
\affiliation{Department of Biological Chemistry and Molecular Pharmacology, Harvard Medical School, Boston, MA, 02115 USA}

\author{Kaifeng Bu}
\email{kfbu@fas.harvard.edu}
\affiliation{Department of Physics, Harvard University, Cambridge, MA 02138, USA}

\author{Liyuan Chen}
\email{liyuanchen@fas.harvard.edu}
\affiliation{Department of Physics, Harvard University, Cambridge, MA 02138, USA}
\affiliation{John A. Paulson School of Engineering and Applied Science, Harvard University, Cambridge, MA 02138, USA}

\author{Haribabu Arthanari}
\email{hari\_arthanari@hms.harvard.edu}
\affiliation{Department of Cancer Biology, Dana-Farber Cancer Institute, Boston, MA, 02215 USA}
\affiliation{Department of Biological Chemistry and Molecular Pharmacology, Harvard Medical School, Boston, MA, 02115 USA}

\author{Arthur Jaffe}
\email{jaffe@g.harvard.edu}
\affiliation{Department of Physics, Harvard University, Cambridge, MA 02138, USA}

\date{\today}

\begin{abstract}
Quantum computers promise to solve computational problems significantly faster than classical computers. These `speed-ups' are achieved by utilizing a resource known as magic. Measuring the amount of magic used by a device allows us to quantify its potential computational power. Without this property, quantum computers are no faster than classical computers. Whether magic can be accurately measured on large-scale quantum computers has remained an open problem. To address this question, we introduce \textit{Pauli instability} as a measure of magic and experimentally measure it on the IBM Eagle quantum processor. We prove that measuring large (i.e., extensive) quantities of magic is intractable. Our results suggest that one may only measure magic when a quantum computer does not provide a speed-up. We support our conclusions with both theoretical and experimental evidence. Our work illustrates the capabilities and limitations of quantum technology in measuring one of the most important resources in quantum computation.
%Magic is one of the most sought-after resources in quantum computation due to its potential in generating quantum computational speed-ups. Quantifying the amount of magic in a quantum device is useful in assessing its potential computational power. We introduce a measure of magic, called the LOTOC. We use this function to show that one can efficiently measure sufficiently small quantities of magic on large-scale quantum computers. We also use the LOTOC to measure magic experimentally on the IBM Eagle quantum processor. We find that measuring larger quantities of magic requires greater measurement precision. This occurs because the number of physical measurements needed grows exponentially with the amount of magic. This implies that measuring extensive quantities of magic is intractable. We conjecture this to be true for any reliable magic measure. Our work illustrates the capabilities and limitations of today's state-of-the-art quantum processors in measuring one of the most important quantum resources.
\end{abstract}

\maketitle

\section{Introduction}

%\rjg{Make more general. Work on abstract and intro. More general sentence on quantum computers. What is magic and why is it needed? Problem statement: it's difficult to measure. Answer: Find that it's still hard to measure. Make it accessible to non-expert. One line about quantum computers. Then why they can solve this. Give an idea of what magic is. For bigger problems, magic becomes a bigger factor. Tells us how many resources we have. We show using both experimental and theoretical resources, this is an intractable problem. Reference the reviewer. We introduce LOTOC, a measure of magic. Direct it to}

Quantum computers~\cite{Feynman1982, LLoyd1996, Ladd2010} are powerful devices with the potential to solve problems in fields as diverse as biology~\cite{Cao2018, Emani2021}, chemistry~\cite{Lanyon2010, Cao2019, RevModPhys.92.015003}, physics~\cite{Feynman1982}, cryptography~\cite{Shor1997}, machine learning~\cite{Biamonte2017}, and finance~\cite{PhysRevA.98.022321}. These devices can perform computations significantly faster than classical computers~\cite{gottesman1998heisenberg}, a feat referred to as quantum advantage~\cite{Shor1997, Bremner_2010, Aaronson2011, Huang2022, Yoganathan_2019, preskill2012quantum, Harrow2017}. In recent years, claims of significant quantum advantages have been made for sampling problems~\cite{Arute_supremacy, Zhong_2020}. 

Quantum computers can only attain such computational speed-ups by utilizing a property known as magic~\cite{Veitch_2014, Howard_2017}. Without this property, quantum computers can perform computations no faster than supercomputers~\cite{gottesman1998heisenberg}. Informally, the amount of magic used by a quantum device quantifies its potential to solve computational problems quickly. Measuring this property is therefore important in assessing the capabilities of real-world quantum computers. However, previous experiments have suggested that such a measurement may be difficult~\cite{Oliviero2022}. In this work, we study whether magic can be measured experimentally on large-scale quantum computers. 

Magic is measured by using functions known as \textit{magic monotones}. Examples of well-known monotones include the robustness of magic~\cite{Howard_2017}, stabilizer rank~\cite{PhysRevX.6.021043}, mana and the relative entropy of magic~\cite{Veitch_2014}, among others~\cite{Seddon_2019, Wang_2019}. These functions have played a central role in finding applications of magic in quantum computation. They have been used to prove bounds on the time needed to classically simulate a computation~\cite{PhysRevLett.115.070501, Bravyi_2016, Howard_2017, Bravyi_2019, Seddon_2019, Seddon_2021, Bu_2019, Bu_2022}, which is useful to identify quantum advantages. Monotones have also been used to bound the cost of generating so-called magic states~\cite{PhysRevA.71.022316,PhysRevLett.124.090505, PhysRevA.86.052329, Haah2017magicstate}; these states are important in realizing an essential feat known as fault-tolerant, universal quantum computation~\cite{Gottesman_1999, Knille2005, PhysRevA.71.022316, Campbell2017}. Furthermore, monotones have been used to link magic to interdisciplinary topics, such as quantum circuit complexity and statistical complexity~\cite{BuPRA19_stat, Bu2022}. 

Our work investigates the hardness of measuring magic monotones in experiments. Monotones are thought to be difficult to measure, as they are typically defined as sums or optimizations over exponentially many variables. This measurement problem has received increasing attention in recent years~\cite{Mi_2021, Oliviero2022, PRXQuantum.4.010301, hamaguchi2023handbook}. In 2021, Google conducted an experiment detecting signatures of magic on their Sycamore quantum processor~\cite{Mi_2021}. In 2023, a magic monotone was introduced to explain this measurement~\cite{Garcia2023Resource}. In 2022, Google's result was followed by a measurement of a new magic monotone on IBM's quantum processor~\cite{Oliviero2022}. This experiment required a large number of physical measurements (exponentially large in the processor size), making it intractable for large systems.

It remained an open question whether other magic monotones could be measured on large-scale quantum computers. In 2023, a measurement of a magic monotone known as the additive Bell magic was made on IonQ's quantum computer~\cite{PRXQuantum.4.010301} and was believed to be tractable at the large scale. This was followed by a measurement of the additive Bell magic on a logical quantum processor~\cite{Bluvstein2024}. We find that further inspection is needed to determine whether one can indeed measure magic on large quantum devices.

In this paper, we introduce a magic monotone, which we call \textit{Pauli instability}, and measure it on the IBM Eagle quantum processor. We use Pauli instability to show that larger quantities of magic are more difficult to measure (Theorem~\ref{Thm:SamplesComplexity}). As the magic increases, the number of physical measurements needed grows exponentially. We show that sufficiently small quantities of magic can be efficiently measured on large-scale quantum computers. We also find that, when magic is extensive, measurement is intractable. We conjecture this to be true for any reliable magic monotone. Our results lead us to posit that one may only measure magic when a quantum computer does not exhibit a quantum advantage. Furthermore, our results showcase the role that chaos plays in measuring magic. Our techniques are compatible with quantum platforms with single-qubit readout~\cite{Bluvstein2022, Arute_supremacy, Wright2019}.

\section{Preliminaries}
%Examples of resources include entanglement~\cite{HorodeckiRMP09, ChitambarRMP19, PhysRevLett.70.1895, Islam2015}, magic~\cite{Veitch_2014, Howard_2017}, and coherence~\cite{aberg2006quantifying, baumgratz2014quantifying,winter2016operational,bu2017maximum,Streltsov2017colloquium, Streltsov_2017,Tajima_universal_2021,Tajima_trade_2022}, among others~\cite{Garcia2023Resource, garcia2024resourcetheorynonrevivalsapplications}. 
We introduce the mathematical definition of magic. We define the set of $n$-qubit Pauli strings as ${Q^{\otimes n}=\{\otimes
_{i=1}^n P^{(i)} : P^{(i)}\in\{I,X,Y,Z\}\}}$,
where $n$ is the system size, $I$ is the identity, and $X,Y,Z$ are Pauli operators. A Clifford unitary, $U$, is defined as a unitary which maps any Pauli string $P$ to another Pauli string $P'$ (up to a phase $\phi$) under conjugation: $U^\dagger P U=e^{-i\phi}P'$. All Clifford unitaries can be generated by the gate set $\{H, S, \mathrm{CNOT}\}$, where $H=\frac{1}{\sqrt{2}}\begin{pmatrix}1 & 1\\1 & -1 \end{pmatrix}$, $S=\mathrm{diag}(1, i)$, and $\mathrm{CNOT}$ denotes the controlled not gate. 

Unitaries which are non-Clifford are defined to contain magic. Intuitively, magic quantifies the distance between a given unitary and the set of Clifford unitaries. The more magic a unitary contains, the more resourceful it is thought to be in completing computational tasks. The best-known example of a non-Clifford gate is the T gate, defined as $T=\mathrm{diag}(1,e^{i\pi/4})$. 

\section{Main Results}
We introduce a monotone, which we call Pauli instability, to measure the magic of a unitary.
\begin{definition}
The Pauli instability of a unitary $U$ is 
\begin{equation}\label{Eq:Pauli_instability}
    \insta(U)=-\log\left[\av{P_1,P_2\in Q^{\otimes n}}\abs{\OTOC(U,P_1,P_2)}\right],
\end{equation}
where $\OTOC(U,P_1,P_2)=\frac{1}{2^n}\tr{ U^\dagger P_1 U P_2 U^\dagger P_1 U P_2}$ and $\mathbb{E}$ denotes the uniform expectation over $Q^{\otimes n}$.
\end{definition}
Here, $\OTOC(U,P_1,P_2)$ is the out-of-time-ordered correlator. Pauli instability satisfies the following properties:
\begin{enumerate}
    \item (Faithfulness) $\insta(U)\geq 0$ for all unitaries $U$ and $\insta(U)=0$ iff $U$ is a Clifford unitary. 
    \item (Invariance) $\insta(V_1UV_2)=\insta(U)$,  for any Clifford unitaries $V_1$ and $V_2$.
    \item (Additivity) $\insta(U_1\otimes U_2)=\insta(U_1)+\insta(U_2)$.
    \item (Scaling with T gates) $\insta(T^{\otimes k}\otimes I^{\otimes n-k})= k\log(4/3)$.
\end{enumerate}
Faithfulness guarantees that Clifford unitaries have no magic, while non-Clifford unitaries have positive magic.  
Invariance guarantees that appending a Clifford unitary to a  circuit cannot increase the measure of magic. These two properties make Pauli instability a resource monotone. Furthermore, additivity implies that for any Clifford unitary, $V$, one has the following invariance relation: ${\insta(U\otimes V)=\insta(U)}$. Scaling with T gates makes Pauli instability a reliable monotone, as the number of T gates in a unitary often gives an indication of its classical simulation cost~\cite{Howard_2017}. This property holds regardless of the position of the T gates~\footnote{Namely, ${\insta (\otimes_{i=1}^nT^{k_{i}})= k\log(4/3)}$, where $k_j\in \{0,1\}$ and $\sum_{j=1}^{n}k_j=k$.}.

Pauli instability can also be interpreted as a measure of chaos, since the OTOC is traditionally used to measure the onset of chaos in chaotic quantum systems~\cite{Fan_2017}. The OTOC notably characterizes a feature of chaos known as scrambling, which describes the delocalization of quantum information. For example, a completely scrambling unitary $U$ can map a Pauli string $P$ to a superposition of many Pauli strings: $U^\dagger P U=\sum_{i}c_iP_i$; this is interpreted as `delocalization' in Pauli space. This is a property of non-Clifford unitaries (this can be seen from the definition of Clifford unitaries). This property can result in $\abs{\OTOC(U,P_1,P_2)}$ taking on values near 0, leading to a positive value of Pauli instability. By contrast, when $U$ is Clifford, $\abs{\OTOC(U,P_1,P_2)}=1$ and Pauli instability is 0. In this way, the monotone exploits the chaotic properties of unitaries to measure their magic. Furthermore, we name the monotone Pauli instability because it quantifies how well a unitary evolves a Pauli string away from itself.

We now introduce an approach to approximate Pauli instability. Due to the average over $Q^{\otimes n}$ in Eq.~\eqref{Eq:Pauli_instability}, an exact measurement of Pauli instability necessitates computing $16^{n}$ terms. Such a measurement is not feasible for large-scale quantum systems. However, one can efficiently approximate Pauli instability by uniformly sampling $N$ pairs of Pauli strings $\{(P_1^{(i)},P_{2}^{(i)})\}_{i=1}^N$ from $Q^{\otimes n}$, where each string in the pair is sampled independently. We refer to $N$ as the \textit{Pauli sample complexity}. By computing the OTOC for each pair of strings, one can construct an approximator of Pauli instability:
\begin{equation}\label{Eq:insta_Unbiased}
    \insta_{N}(U)=-\log\left[\frac{1}{N}\sum_{i=1}^{N}\abs{\OTOC(U,P_1^{(i)},P_2^{(i)})}\right].
\end{equation}
The finite sampling over $Q^{\otimes n}$ introduces an error of approximation. One can use Hoeffding's inequality to compute the Pauli sample complexity needed to approximate Pauli instability up to a given error.

%\ajq{The following is an alternative statement of Theorem 2. What do you think?} \rjg{I prefer the first version of this theorem because the equation is more readable. }
%\setcounter{theorem}{\thetheorem-1}
%\begin{theorem}[Pauli sample complexity]\label{Thm:SampleComplexity}
%Given $\epsilon,\delta>0$, take   $N\geqslant\frac{\ln(1/\delta)}{2\epsilon^2}$. Then  
%\begin{equation}
%   {\rm Prob}\left \{ \abs{e^{-\LOTOC_N(U)}-e^{-\LOTOC(U)}}< \epsilon \right\} >1-\delta\;.
%\end{equation}
%\end{theorem}

\begin{theorem}[Pauli sample complexity]\label{Thm:SamplesComplexity}
Let $\delta, \eta>0$. Then $\abs{\insta_N(U)-\insta(U)}<\eta$ with probability at least $1-\delta$ when the Pauli sample complexity is
\begin{equation}
    N=e^{2\insta(U)}f(\eta, \delta).
\end{equation}
Here, $f(\eta,\delta)=\frac{\ln(1/\delta)}{2(1-e^{g\eta})^2}$ and $g=\mathrm{sign}(\insta(U)-\insta_N(U))$.
\end{theorem}
Intuitively, Theorem~\ref{Thm:SamplesComplexity} shows that measuring more magic requires exponentially more samples. This theorem can help us identify the regime where one can accurately and efficiently measure magic. Here, `efficiently' means that the Pauli sample complexity satisfies $N=poly(n)$. We say that a measurement is intractable when $N=exp(n)$. `Accurately' means that $\eta$ satisfies $\eta=\gamma \insta(U)$, where $0<\gamma<1$ is a constant factor. 

\begin{corollary}\label{Cor_insta}
Magic can be efficiently and accurately approximated when $\insta(U)=\log(n)$. However, an accurate approximation is intractable when $\insta(U)=linear(n)$.
\end{corollary}

\begin{figure*}[t!]
\begin{center}
\scalebox{.44}{
\begin{tikzpicture}
\renewcommand{\xglobalshift}{0*\xrel}
\renewcommand{\yglobalshift}{0*\yrel}
    
	\renewcommand{\nodenum}{v9}
    \renewcommand{\name}{\includegraphics[scale=.65]{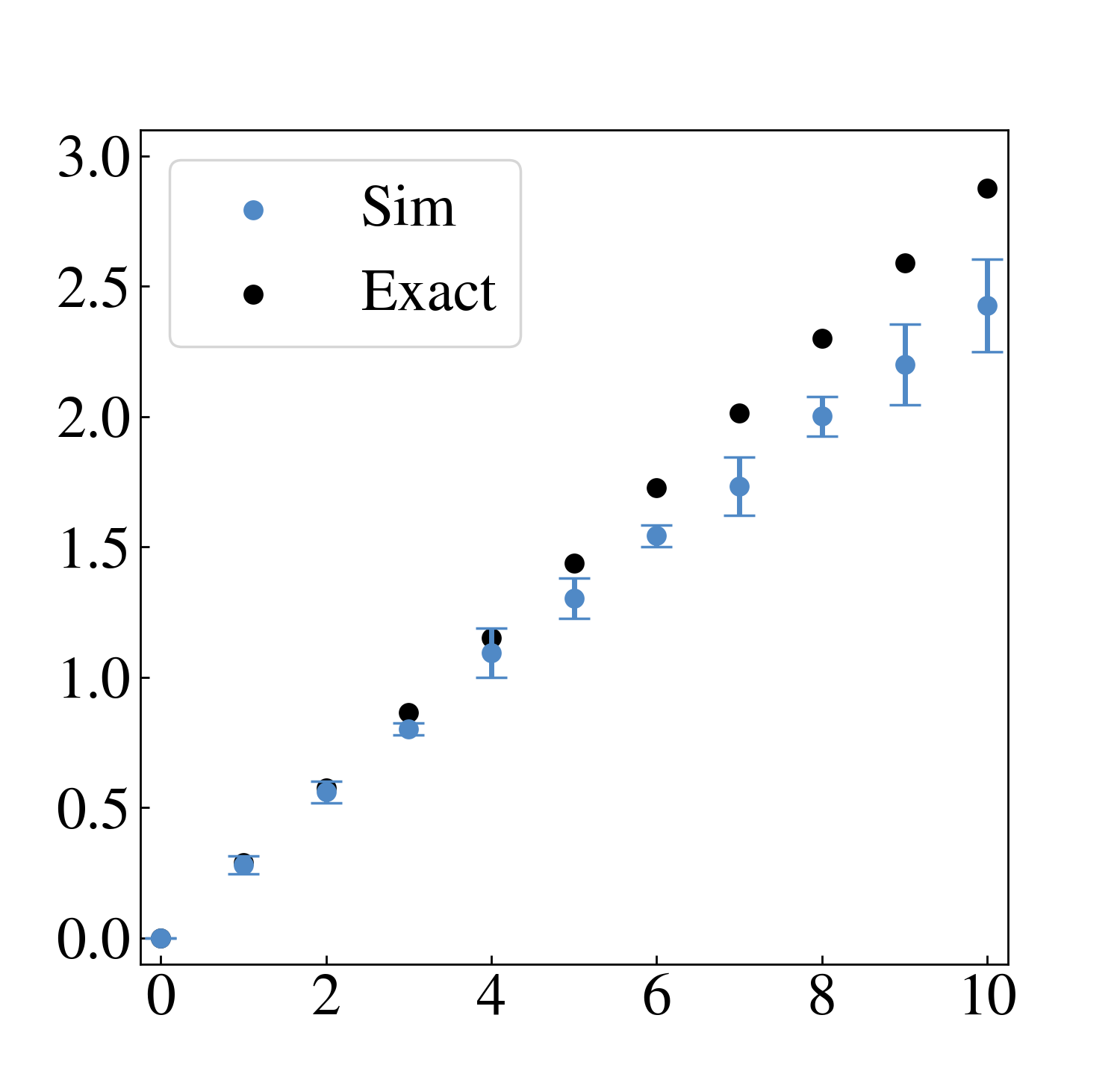}}
	\renewcommand{\xpos}{-8.5*\xrel+\xglobalshift}
    \renewcommand{\ypos}{3*\rowspace+2.4*\yrel+\yglobalshift}
    \renewcommand{\height}{10*\heightsingle}
    \renewcommand{\width}{1.5*\widthsingle}
    \node[] (\nodenum) at (\xpos,\ypos) {\name};

	\renewcommand{\nodenum}{v9}
    \renewcommand{\name}{\includegraphics[scale=.65]{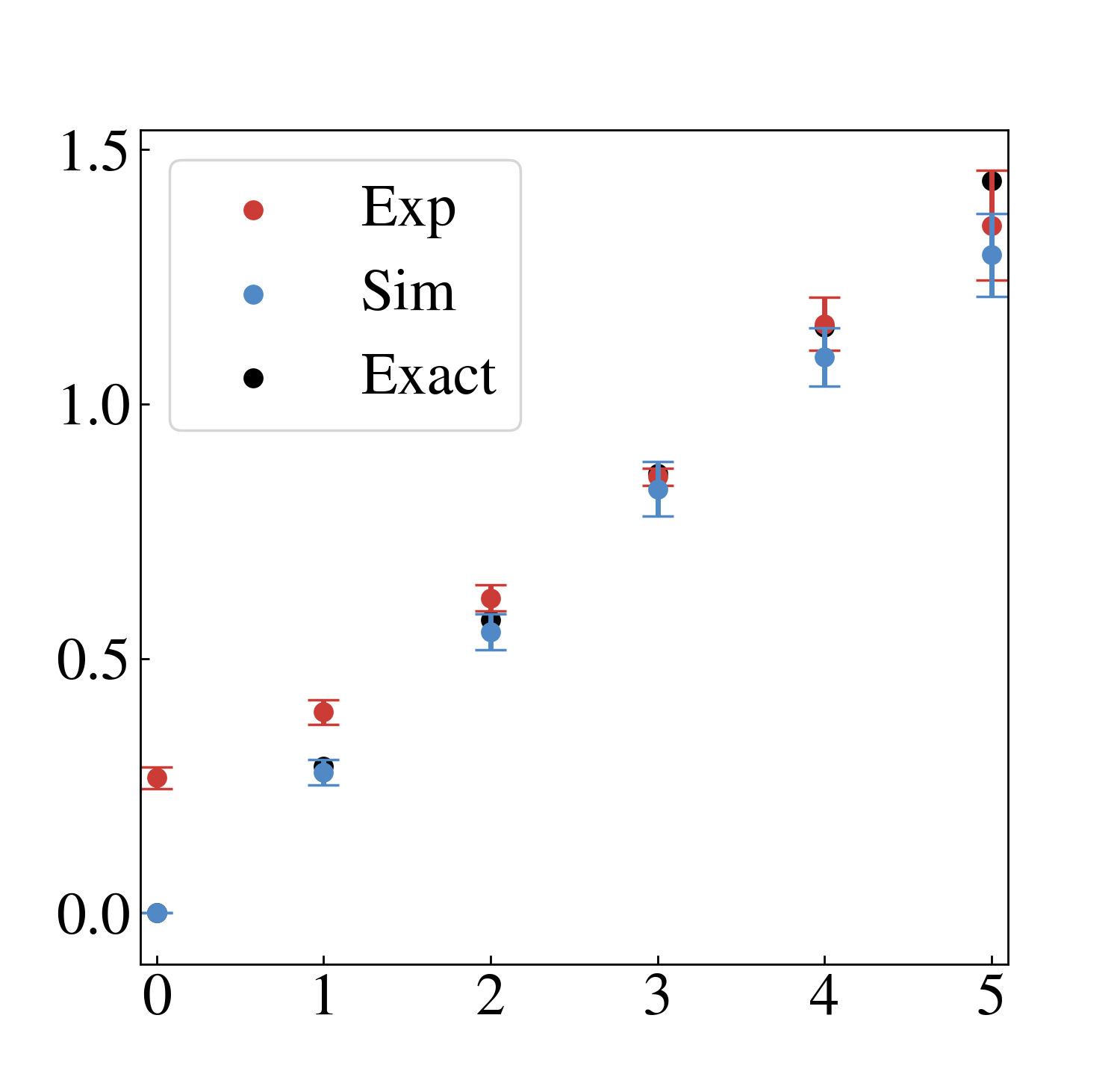}}
	\renewcommand{\xpos}{.75*\xrel+\xglobalshift}
    \renewcommand{\ypos}{3.5*\rowspace+1.4*\yrel+\yglobalshift}
    \renewcommand{\height}{10*\heightsingle}
    \renewcommand{\width}{1.5*\widthsingle}
    \node[] (\nodenum) at (\xpos,\ypos) {\name};

    \renewcommand{\nodenum}{v9}
    \renewcommand{\name}{\includegraphics[scale=.65]{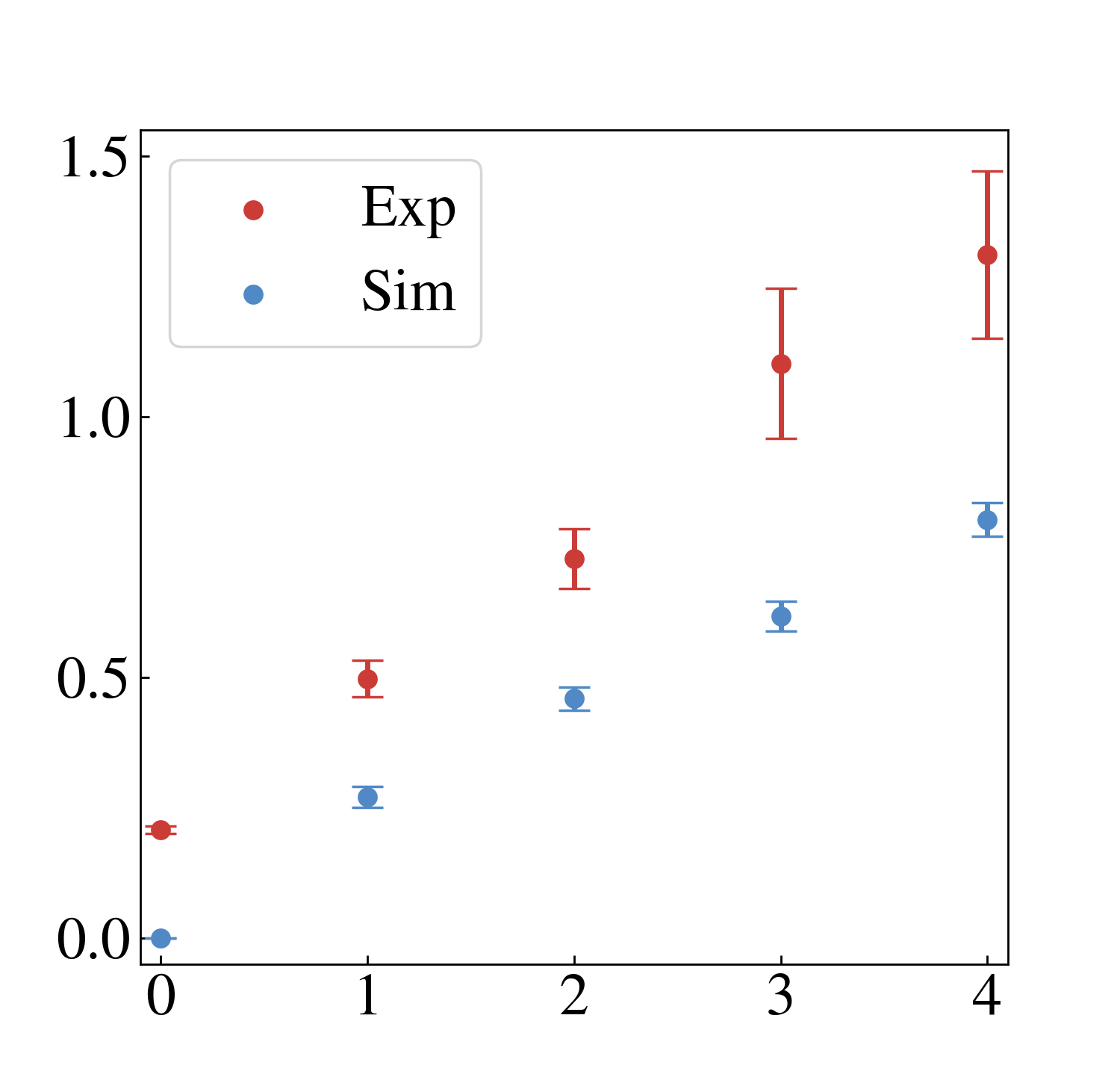}}
	\renewcommand{\xpos}{23*\xrel+\xglobalshift}
    \renewcommand{\ypos}{3.5*\rowspace+1.4*\yrel+\yglobalshift}
    \renewcommand{\height}{10*\heightsingle}
    \renewcommand{\width}{1.5*\widthsingle}
    \node[] (\nodenum) at (\xpos,\ypos) {\name};

    \scalebox{2}{
	\renewcommand{\nodenum}{v9}
    \renewcommand{\name}{T gates}
	\renewcommand{\xpos}{-4.2*\xrel+\xglobalshift}
    \renewcommand{\ypos}{-.3*\rowspace+\yglobalshift}
    \renewcommand{\height}{2.5*\heightsingle}
    \renewcommand{\width}{1.5*\widthsingle}
    \node[] (\nodenum) at (\xpos,\ypos) {\name};
    }
    
    \scalebox{2}{
	\renewcommand{\nodenum}{v9}
    \renewcommand{\name}{T gates}
	\renewcommand{\xpos}{.5*\xrel+\xglobalshift}
    \renewcommand{\ypos}{-.3*\rowspace+\yglobalshift}
    \renewcommand{\height}{2.5*\heightsingle}
    \renewcommand{\width}{1.5*\widthsingle}
    \node[] (\nodenum) at (\xpos,\ypos) {\name};
    }
    \scalebox{2}{
	\renewcommand{\nodenum}{v9}
    \renewcommand{\name}{T gates}
	\renewcommand{\xpos}{11.7*\xrel+\xglobalshift}
    \renewcommand{\ypos}{-.3*\rowspace+\yglobalshift}
    \renewcommand{\height}{2.5*\heightsingle}
    \renewcommand{\width}{1.5*\widthsingle}
    \node[] (\nodenum) at (\xpos,\ypos) {\name};
    }
    \scalebox{2}{
	\renewcommand{\nodenum}{v9}
    \renewcommand{\name}{Magic}
	\renewcommand{\xpos}{-6.75*\xrel+\xglobalshift}
    \renewcommand{\ypos}{2*\rowspace+\yglobalshift}
    \renewcommand{\height}{2.5*\heightsingle}
    \renewcommand{\width}{1.5*\widthsingle}
    \node[rotate=90] (\nodenum) at (\xpos,\ypos) {\name};
    }
    \scalebox{2}{
	\renewcommand{\nodenum}{v9}
    \renewcommand{\name}{Magic}
	\renewcommand{\xpos}{-2*\xrel+\xglobalshift}
    \renewcommand{\ypos}{2*\rowspace+\yglobalshift}
    \renewcommand{\height}{2.5*\heightsingle}
    \renewcommand{\width}{1.5*\widthsingle}
    \node[rotate=90] (\nodenum) at (\xpos,\ypos) {\name};
    }    
    \scalebox{2}{
	\renewcommand{\nodenum}{v9}
    \renewcommand{\name}{Magic}
	\renewcommand{\xpos}{9.1*\xrel+\xglobalshift}
    \renewcommand{\ypos}{2*\rowspace+\yglobalshift}
    \renewcommand{\height}{2.5*\heightsingle}
    \renewcommand{\width}{1.5*\widthsingle}
    \node[rotate=90] (\nodenum) at (\xpos,\ypos) {\name};
    }    

    \scalebox{2}{
	\renewcommand{\nodenum}{v9}
    \renewcommand{\name}{(a)}
	\renewcommand{\xpos}{-4.2*\xrel+\xglobalshift}
    \renewcommand{\ypos}{-.8*\rowspace+\yglobalshift}
    \renewcommand{\height}{2.5*\heightsingle}
    \renewcommand{\width}{1.5*\widthsingle}
    \node[] (\nodenum) at (\xpos,\ypos) {\name};
    }
    
    \scalebox{2}{
	\renewcommand{\nodenum}{v9}
    \renewcommand{\name}{(b)}
	\renewcommand{\xpos}{.5*\xrel+\xglobalshift}
    \renewcommand{\ypos}{-.8*\rowspace+\yglobalshift}
    \renewcommand{\height}{2.5*\heightsingle}
    \renewcommand{\width}{1.5*\widthsingle}
    \node[] (\nodenum) at (\xpos,\ypos) {\name};
    }
    \scalebox{2}{
	\renewcommand{\nodenum}{v9}
    \renewcommand{\name}{(c)}
	\renewcommand{\xpos}{6.5*\xrel+\xglobalshift}
    \renewcommand{\ypos}{-.8*\rowspace+\yglobalshift}
    \renewcommand{\height}{2.5*\heightsingle}
    \renewcommand{\width}{1.5*\widthsingle}
    \node[] (\nodenum) at (\xpos,\ypos) {\name};
    }
    \scalebox{2}{
	\renewcommand{\nodenum}{v9}
    \renewcommand{\name}{(d)}
	\renewcommand{\xpos}{11.7*\xrel+\xglobalshift}
    \renewcommand{\ypos}{-.8*\rowspace+\yglobalshift}
    \renewcommand{\height}{2.5*\heightsingle}
    \renewcommand{\width}{1.5*\widthsingle}
    \node[] (\nodenum) at (\xpos,\ypos) {\name};
    }

    %U_k gate
    \draw [thick,color=charcoal]
  (-7*\xrel+\xglobalshift,2.1*\rowspace+\yrel+\yglobalshift)--++(2*\xrel,0)
    (-7*\xrel+\xglobalshift,2.1*\rowspace-\yrel+\yglobalshift)--++(2*\xrel,0);
    
	\renewcommand{\nodenum}{v9}
    \renewcommand{\name}{\Large $U_k$}
	\renewcommand{\xpos}{-6*\xrel+\xglobalshift}
    \renewcommand{\ypos}{2.1*\rowspace+\yglobalshift}
    \renewcommand{\height}{2.5*\heightsingle}
    \renewcommand{\width}{1.5*\widthsingle}
    \node[rectangle, fill=mustard,  line width =.3mm,rounded corners, minimum width=\width, minimum height= \height, draw=charcoal] (\nodenum) at (\xpos,\ypos) {\name};

    %U_k gate
    \draw [thick,color=charcoal]    (2*\xrel+\xglobalshift,2.1*\rowspace+\yrel+\yglobalshift)--++(2*\xrel,0)
    (2*\xrel+\xglobalshift,2.1*\rowspace-\yrel+\yglobalshift)--++(2*\xrel,0);
    
    \renewcommand{\nodenum}{v9}
    \renewcommand{\name}{\Large $U_k$}
	\renewcommand{\xpos}{3*\xrel+\xglobalshift}
    \renewcommand{\ypos}{2.1*\rowspace+\yglobalshift}
    \renewcommand{\height}{2.5*\heightsingle}
    \renewcommand{\width}{1.5*\widthsingle}
    \node[rectangle, fill=mustard,  line width =.3mm,rounded corners, minimum width=\width, minimum height= \height, draw=charcoal] (\nodenum) at (\xpos,\ypos) {\name};

    %U_k gate
    \draw [thick,color=charcoal]    (24.4*\xrel+\xglobalshift,2.1*\rowspace+\yrel+\yglobalshift)--++(2*\xrel,0)
    (24.4*\xrel+\xglobalshift,2.2*\rowspace-\yrel+\yglobalshift)--++(2*\xrel,0);
    
    \renewcommand{\nodenum}{v9}
    \renewcommand{\name}{\Large $V_k$}
	\renewcommand{\xpos}{25.4*\xrel+\xglobalshift}
    \renewcommand{\ypos}{2.1*\rowspace+\yglobalshift}
    \renewcommand{\height}{2.5*\heightsingle}
    \renewcommand{\width}{1.5*\widthsingle}
    \node[rectangle, fill=pastelorange,  line width =.3mm,rounded corners, minimum width=\width, minimum height= \height, draw=charcoal] (\nodenum) at (\xpos,\ypos) {\name};

%First diagram
\renewcommand{\xglobalshift}{6*\xrel}
\renewcommand{\yglobalshift}{5*\rowspace}

    \draw [thick,color=charcoal]
    (2*\xrel+\xglobalshift,\yrel+2*\rowspace+\yglobalshift)--++(2*\xrel,0)
    (2*\xrel+\xglobalshift,\yrel+\rowspace+\yglobalshift)--++(2*\xrel,0)
    (2*\xrel+\xglobalshift,\yrel+\yglobalshift)--++(2*\xrel,0)
    (2*\xrel+\xglobalshift,\yrel-\rowspace+\yglobalshift)--++(2*\xrel,0);

    \draw [thick,color=charcoal]
    (-.75*\xrel+\xglobalshift,\yrel+2*\rowspace+\yglobalshift)--++(2*\xrel,0)
    (-.75*\xrel+\xglobalshift,\yrel+\rowspace+\yglobalshift)--++(2*\xrel,0)
    (-.75*\xrel+\xglobalshift,\yrel+\yglobalshift)--++(2*\xrel,0)
    (-.75*\xrel+\xglobalshift,\yrel-\rowspace+\yglobalshift)--++(2*\xrel,0);
    
	\renewcommand{\nodenum}{v9}
    \renewcommand{\name}{\Large $U_k$}
	\renewcommand{\xpos}{0.25*\xrel+\xglobalshift}
    \renewcommand{\ypos}{\rowspace+\yglobalshift}
    \renewcommand{\height}{6*\heightsingle}
    \renewcommand{\width}{1.5*\widthsingle}
    \node[rectangle, fill=mustard,  line width =.3mm,rounded corners, minimum width=\width, minimum height= \height, draw=charcoal] (\nodenum) at (\xpos,\ypos) {\name};

	\renewcommand{\nodenum}{v9}
    \renewcommand{\name}{\Large $=$}
	\renewcommand{\xpos}{1.65*\xrel+\xglobalshift}
    \renewcommand{\ypos}{1*\rowspace+\yglobalshift}
    \renewcommand{\height}{4*\heightsingle}
    \renewcommand{\width}{1.5*\widthsingle}
    \node[] (\nodenum) at (\xpos,\ypos) {\name};
    
\renewcommand{\xglobalshift}{9*\xrel}
\renewcommand{\yglobalshift}{5*\rowspace}
    
    \renewcommand{\nodenum}{v1}
    \renewcommand{\name}{$T$}
	\renewcommand{\xpos}{0*\xrel+\xglobalshift}
    \renewcommand{\ypos}{2*\rowspace+\yrel+\yglobalshift}
    \renewcommand{\height}{\heightsingle}
    \renewcommand{\width}{\widthsingle}
    \node[rectangle, fill=evergreen,  line width =.3mm,rounded corners, minimum width=\width, minimum height= \height, draw=charcoal] (\nodenum) at (\xpos,\ypos) {\name};

    \renewcommand{\nodenum}{v1}
    \renewcommand{\name}{$T$}
	\renewcommand{\xpos}{0*\xrel+\xglobalshift}
    \renewcommand{\ypos}{1*\rowspace+\yrel+\yglobalshift}
    \renewcommand{\height}{\heightsingle}
    \renewcommand{\width}{\widthsingle}
    \node[rectangle, fill=evergreen,  line width =.3mm,rounded corners, minimum width=\width, minimum height= \height, draw=charcoal] (\nodenum) at (\xpos,\ypos) {\name};

    \renewcommand{\xpos}{0*\xrel}
	\draw [decorate,line width=.75pt,color=charcoal,decoration={brace,amplitude=5pt},xshift=\xrel,yshift=-\rowspace]	(1.25*\xrel+\xglobalshift,2*\rowspace+\yrel+\yglobalshift) --++ (0,-1*\rowspace) node [black,midway,xshift=9pt] {};
 
    \renewcommand{\nodenum}{v1}
    \renewcommand{\name}{\Large $k$}
	\renewcommand{\xpos}{1.75*\xrel+\xglobalshift}
    \renewcommand{\ypos}{2*\rowspace+\yglobalshift}
    \renewcommand{\height}{\heightsingle}
    \renewcommand{\width}{\widthsingle}
    \node[] (\nodenum) at (\xpos,\ypos) {\name};
    
%Second diagram
\renewcommand{\xglobalshift}{6*\xrel}
\renewcommand{\yglobalshift}{\yrel}   
    
    \draw [thick,color=charcoal]
    (2*\xrel+\xglobalshift,\yrel+2*\rowspace+\yglobalshift)--++(10*\xrel,0)
    (2*\xrel+\xglobalshift,\yrel+\rowspace+\yglobalshift)--++(10*\xrel,0)
    (2*\xrel+\xglobalshift,\yrel+\yglobalshift)--++(10*\xrel,0)
    (2*\xrel+\xglobalshift,\yrel-\rowspace+\yglobalshift)--++(10*\xrel,0);

    \draw [thick,color=charcoal]
    (-.75*\xrel+\xglobalshift,\yrel+2*\rowspace+\yglobalshift)--++(2*\xrel,0)
    (-.75*\xrel+\xglobalshift,\yrel+\rowspace+\yglobalshift)--++(2*\xrel,0)
    (-.75*\xrel+\xglobalshift,\yrel+\yglobalshift)--++(2*\xrel,0)
    (-.75*\xrel+\xglobalshift,\yrel-\rowspace+\yglobalshift)--++(2*\xrel,0);
    
	\renewcommand{\nodenum}{v9}
    \renewcommand{\name}{\Large $V_k$}
	\renewcommand{\xpos}{0.25*\xrel+\xglobalshift}
    \renewcommand{\ypos}{\rowspace+\yglobalshift}
    \renewcommand{\height}{6*\heightsingle}
    \renewcommand{\width}{1.5*\widthsingle}
    \node[rectangle, fill=pastelorange,  line width =.3mm,rounded corners, minimum width=\width, minimum height= \height, draw=charcoal] (\nodenum) at (\xpos,\ypos) {\name};

	\renewcommand{\nodenum}{v9}
    \renewcommand{\name}{\Large $=$}
	\renewcommand{\xpos}{1.65*\xrel+\xglobalshift}
    \renewcommand{\ypos}{\rowspace+\yglobalshift}
    \renewcommand{\height}{4*\heightsingle}
    \renewcommand{\width}{1.5*\widthsingle}
    \node[] (\nodenum) at (\xpos,\ypos) {\name};
    
\renewcommand{\xglobalshift}{9*\xrel}
\renewcommand{\yglobalshift}{\yrel}
    
    \renewcommand{\nodenum}{v1}
    \renewcommand{\name}{$H$}
	\renewcommand{\xpos}{0*\xrel+\xglobalshift}
    \renewcommand{\ypos}{2*\rowspace+\yrel+\yglobalshift}
    \renewcommand{\height}{\heightsingle}
    \renewcommand{\width}{\widthsingle}
    \node[rectangle, fill=egg,  line width =.3mm,rounded corners, minimum width=\width, minimum height= \height, draw=charcoal] (\nodenum) at (\xpos,\ypos) {\name};

    \renewcommand{\nodenum}{v1}
    \renewcommand{\name}{$H$}
	\renewcommand{\xpos}{0*\xrel+\xglobalshift}
    \renewcommand{\ypos}{1*\rowspace+\yrel+\yglobalshift}
    \renewcommand{\height}{\heightsingle}
    \renewcommand{\width}{\widthsingle}
    \node[rectangle, fill=egg,  line width =.3mm,rounded corners, minimum width=\width, minimum height= \height, draw=charcoal] (\nodenum) at (\xpos,\ypos) {\name};
    
    \renewcommand{\nodenum}{v3}
    \renewcommand{\name}{$H$}
	\renewcommand{\xpos}{0*\xrel+\xglobalshift}
    \renewcommand{\ypos}{\yrel+\yglobalshift}
    \renewcommand{\height}{\heightsingle}
    \renewcommand{\width}{\widthsingle}
    \node[rectangle, fill=egg,  line width =.3mm,rounded corners, minimum width=\width, minimum height= \height, draw=charcoal] (\nodenum) at (\xpos,\ypos) {\name};

    \renewcommand{\nodenum}{v3}
    \renewcommand{\name}{$H$}
	\renewcommand{\xpos}{0*\xrel+\xglobalshift}
    \renewcommand{\ypos}{\yrel-\rowspace+\yglobalshift}
    \renewcommand{\height}{\heightsingle}
    \renewcommand{\width}{\widthsingle}
    \node[rectangle, fill=egg,  line width =.3mm,rounded corners, minimum width=\width, minimum height= \height, draw=charcoal] (\nodenum) at (\xpos,\ypos) {\name};
    
    %CNOT gates
    \renewcommand{\xpos}{1*\xrel+\xglobalshift}
    \renewcommand{\ypos}{2*\rowspace+\yrel+\yglobalshift}
    \renewcommand{\height}{.2*\heightsingle}
    \node[circle, fill=charcoal, line width =.3mm, minimum height=\height,  draw=charcoal] (\nodenum) at (\xpos,\ypos) {};

    \draw [thick,color=charcoal]
    (1*\xrel+\xglobalshift, 2*\rowspace+\yrel+\yglobalshift)--++(0,-2.3*\yrel);
    
	\renewcommand{\xpos}{1*\xrel+\xglobalshift}
    \renewcommand{\ypos}{1*\rowspace+\yrel+\yglobalshift}
    \renewcommand{\height}{.2*\heightsingle}
    \node[circle,  line width =.3mm, minimum height=\height,  draw=charcoal] (\nodenum) at (\xpos,\ypos) {};

    %CNOT gates
    \renewcommand{\xpos}{1*\xrel+\xglobalshift}
    \renewcommand{\ypos}{0*\rowspace+\yrel+\yglobalshift}
    \renewcommand{\height}{.2*\heightsingle}
    \node[circle, fill=charcoal, line width =.3mm, minimum height=\height,  draw=charcoal] (\nodenum) at (\xpos,\ypos) {};

    \draw [thick,color=charcoal]
    (1*\xrel+\xglobalshift, 0*\rowspace+\yrel+\yglobalshift)--++(0,-2.3*\yrel);
    
	\renewcommand{\xpos}{1*\xrel+\xglobalshift}
    \renewcommand{\ypos}{-1*\rowspace+\yrel+\yglobalshift}
    \renewcommand{\height}{.2*\heightsingle}
    \node[circle,  line width =.3mm, minimum height=\height,  draw=charcoal] (\nodenum) at (\xpos,\ypos) {};
    
    %CNOT gates
    \renewcommand{\xpos}{1.75*\xrel+\xglobalshift}
    \renewcommand{\ypos}{1*\rowspace+\yrel+\yglobalshift}
    \renewcommand{\height}{.2*\heightsingle}
    \node[circle, fill=charcoal, line width =.3mm, minimum height=\height,  draw=charcoal] (\nodenum) at (\xpos,\ypos) {};

    \draw [thick,color=charcoal]
    (1.75*\xrel+\xglobalshift, 1*\rowspace+\yrel+\yglobalshift)--++(0,-2.3*\yrel);
    
	\renewcommand{\xpos}{1.75*\xrel+\xglobalshift}
    \renewcommand{\ypos}{0*\rowspace+\yrel+\yglobalshift}
    \renewcommand{\height}{.2*\heightsingle}
    \node[circle,  line width =.3mm, minimum height=\height,  draw=charcoal] (\nodenum) at (\xpos,\ypos) {};	
    
    \renewcommand{\nodenum}{v3}
    \renewcommand{\name}{$S$}
	\renewcommand{\xpos}{2.5*\xrel+\xglobalshift}
    \renewcommand{\ypos}{\yrel+2*\rowspace+\yglobalshift}
    \renewcommand{\height}{\heightsingle}
    \renewcommand{\width}{\widthsingle}
    \node[rectangle, fill=egg,  line width =.3mm,rounded corners, minimum width=\width, minimum height= \height, draw=charcoal] (\nodenum) at (\xpos,\ypos) {\name};
    
    \renewcommand{\nodenum}{v3}
    \renewcommand{\name}{$S$}
	\renewcommand{\xpos}{2.5*\xrel+\xglobalshift}
    \renewcommand{\ypos}{\yrel+\rowspace+\yglobalshift}
    \renewcommand{\height}{\heightsingle}
    \renewcommand{\width}{\widthsingle}
    \node[rectangle, fill=egg,  line width =.3mm,rounded corners, minimum width=\width, minimum height= \height, draw=charcoal] (\nodenum) at (\xpos,\ypos) {\name};

    \renewcommand{\nodenum}{v3}
    \renewcommand{\name}{$S$}
	\renewcommand{\xpos}{2.5*\xrel+\xglobalshift}
    \renewcommand{\ypos}{\yrel+\yglobalshift}
    \renewcommand{\height}{\heightsingle}
    \renewcommand{\width}{\widthsingle}
    \node[rectangle, fill=egg,  line width =.3mm,rounded corners, minimum width=\width, minimum height= \height, draw=charcoal] (\nodenum) at (\xpos,\ypos) {\name};

    \renewcommand{\nodenum}{v3}
    \renewcommand{\name}{$S$}
	\renewcommand{\xpos}{2.5*\xrel+\xglobalshift}
    \renewcommand{\ypos}{\yrel-\rowspace+\yglobalshift}
    \renewcommand{\height}{\heightsingle}
    \renewcommand{\width}{\widthsingle}
    \node[rectangle, fill=egg,  line width =.3mm,rounded corners, minimum width=\width, minimum height= \height, draw=charcoal] (\nodenum) at (\xpos,\ypos) {\name};
    
    \renewcommand{\nodenum}{v1}
    \renewcommand{\name}{$T$}
	\renewcommand{\xpos}{3.5*\xrel+\xglobalshift}
    \renewcommand{\ypos}{\yrel+2*\rowspace+\yglobalshift}
    \renewcommand{\height}{\heightsingle}
    \renewcommand{\width}{\widthsingle}
    \node[rectangle, fill=evergreen,  line width =.3mm,rounded corners, minimum width=\width, minimum height= \height, draw=charcoal] (\nodenum) at (\xpos,\ypos) {\name};
    
\renewcommand{\xglobalshift}{13.5*\xrel}
\renewcommand{\yglobalshift}{\yrel}
    
    \renewcommand{\nodenum}{v1}
    \renewcommand{\name}{$H$}
	\renewcommand{\xpos}{0*\xrel+\xglobalshift}
    \renewcommand{\ypos}{2*\rowspace+\yrel+\yglobalshift}
    \renewcommand{\height}{\heightsingle}
    \renewcommand{\width}{\widthsingle}
    \node[rectangle, fill=egg,  line width =.3mm,rounded corners, minimum width=\width, minimum height= \height, draw=charcoal] (\nodenum) at (\xpos,\ypos) {\name};

    \renewcommand{\nodenum}{v1}
    \renewcommand{\name}{$H$}
	\renewcommand{\xpos}{0*\xrel+\xglobalshift}
    \renewcommand{\ypos}{1*\rowspace+\yrel+\yglobalshift}
    \renewcommand{\height}{\heightsingle}
    \renewcommand{\width}{\widthsingle}
    \node[rectangle, fill=egg,  line width =.3mm,rounded corners, minimum width=\width, minimum height= \height, draw=charcoal] (\nodenum) at (\xpos,\ypos) {\name};
    
    \renewcommand{\nodenum}{v3}
    \renewcommand{\name}{$H$}
	\renewcommand{\xpos}{0*\xrel+\xglobalshift}
    \renewcommand{\ypos}{\yrel+\yglobalshift}
    \renewcommand{\height}{\heightsingle}
    \renewcommand{\width}{\widthsingle}
    \node[rectangle, fill=egg,  line width =.3mm,rounded corners, minimum width=\width, minimum height= \height, draw=charcoal] (\nodenum) at (\xpos,\ypos) {\name};

    \renewcommand{\nodenum}{v3}
    \renewcommand{\name}{$H$}
	\renewcommand{\xpos}{0*\xrel+\xglobalshift}
    \renewcommand{\ypos}{\yrel-\rowspace+\yglobalshift}
    \renewcommand{\height}{\heightsingle}
    \renewcommand{\width}{\widthsingle}
    \node[rectangle, fill=egg,  line width =.3mm,rounded corners, minimum width=\width, minimum height= \height, draw=charcoal] (\nodenum) at (\xpos,\ypos) {\name};
    
    %CNOT gates
    \renewcommand{\xpos}{1*\xrel+\xglobalshift}
    \renewcommand{\ypos}{2*\rowspace+\yrel+\yglobalshift}
    \renewcommand{\height}{.2*\heightsingle}
    \node[circle, fill=charcoal, line width =.3mm, minimum height=\height,  draw=charcoal] (\nodenum) at (\xpos,\ypos) {};

    \draw [thick,color=charcoal]
    (1*\xrel+\xglobalshift, 2*\rowspace+\yrel+\yglobalshift)--++(0,-2.3*\yrel);
    
	\renewcommand{\xpos}{1*\xrel+\xglobalshift}
    \renewcommand{\ypos}{1*\rowspace+\yrel+\yglobalshift}
    \renewcommand{\height}{.2*\heightsingle}
    \node[circle,  line width =.3mm, minimum height=\height,  draw=charcoal] (\nodenum) at (\xpos,\ypos) {};

    %CNOT gates
    \renewcommand{\xpos}{1*\xrel+\xglobalshift}
    \renewcommand{\ypos}{0*\rowspace+\yrel+\yglobalshift}
    \renewcommand{\height}{.2*\heightsingle}
    \node[circle, fill=charcoal, line width =.3mm, minimum height=\height,  draw=charcoal] (\nodenum) at (\xpos,\ypos) {};

    \draw [thick,color=charcoal]
    (1*\xrel+\xglobalshift, 0*\rowspace+\yrel+\yglobalshift)--++(0,-2.3*\yrel);
    
	\renewcommand{\xpos}{1*\xrel+\xglobalshift}
    \renewcommand{\ypos}{-1*\rowspace+\yrel+\yglobalshift}
    \renewcommand{\height}{.2*\heightsingle}
    \node[circle,  line width =.3mm, minimum height=\height,  draw=charcoal] (\nodenum) at (\xpos,\ypos) {};
    
    %CNOT gates
    \renewcommand{\xpos}{1.75*\xrel+\xglobalshift}
    \renewcommand{\ypos}{1*\rowspace+\yrel+\yglobalshift}
    \renewcommand{\height}{.2*\heightsingle}
    \node[circle, fill=charcoal, line width =.3mm, minimum height=\height,  draw=charcoal] (\nodenum) at (\xpos,\ypos) {};

    \draw [thick,color=charcoal]
    (1.75*\xrel+\xglobalshift, 1*\rowspace+\yrel+\yglobalshift)--++(0,-2.3*\yrel);
    
	\renewcommand{\xpos}{1.75*\xrel+\xglobalshift}
    \renewcommand{\ypos}{0*\rowspace+\yrel+\yglobalshift}
    \renewcommand{\height}{.2*\heightsingle}
    \node[circle,  line width =.3mm, minimum height=\height,  draw=charcoal] (\nodenum) at (\xpos,\ypos) {};	

    \renewcommand{\nodenum}{v3}
    \renewcommand{\name}{$S$}
	\renewcommand{\xpos}{2.5*\xrel+\xglobalshift}
    \renewcommand{\ypos}{\yrel+2*\rowspace+\yglobalshift}
    \renewcommand{\height}{\heightsingle}
    \renewcommand{\width}{\widthsingle}
    \node[rectangle, fill=egg,  line width =.3mm,rounded corners, minimum width=\width, minimum height= \height, draw=charcoal] (\nodenum) at (\xpos,\ypos) {\name};
    
    \renewcommand{\nodenum}{v3}
    \renewcommand{\name}{$S$}
	\renewcommand{\xpos}{2.5*\xrel+\xglobalshift}
    \renewcommand{\ypos}{\yrel+\rowspace+\yglobalshift}
    \renewcommand{\height}{\heightsingle}
    \renewcommand{\width}{\widthsingle}
    \node[rectangle, fill=egg,  line width =.3mm,rounded corners, minimum width=\width, minimum height= \height, draw=charcoal] (\nodenum) at (\xpos,\ypos) {\name};

    \renewcommand{\nodenum}{v3}
    \renewcommand{\name}{$S$}
	\renewcommand{\xpos}{2.5*\xrel+\xglobalshift}
    \renewcommand{\ypos}{\yrel+\yglobalshift}
    \renewcommand{\height}{\heightsingle}
    \renewcommand{\width}{\widthsingle}
    \node[rectangle, fill=egg,  line width =.3mm,rounded corners, minimum width=\width, minimum height= \height, draw=charcoal] (\nodenum) at (\xpos,\ypos) {\name};

    \renewcommand{\nodenum}{v3}
    \renewcommand{\name}{$S$}
	\renewcommand{\xpos}{2.5*\xrel+\xglobalshift}
    \renewcommand{\ypos}{\yrel-\rowspace+\yglobalshift}
    \renewcommand{\height}{\heightsingle}
    \renewcommand{\width}{\widthsingle}
    \node[rectangle, fill=egg,  line width =.3mm,rounded corners, minimum width=\width, minimum height= \height, draw=charcoal] (\nodenum) at (\xpos,\ypos) {\name};
    
    \renewcommand{\nodenum}{v1}
    \renewcommand{\name}{$T$}
	\renewcommand{\xpos}{3.5*\xrel+\xglobalshift}
    \renewcommand{\ypos}{\yrel+1*\rowspace+\yglobalshift}
    \renewcommand{\height}{\heightsingle}
    \renewcommand{\width}{\widthsingle}
    \node[rectangle, fill=evergreen,  line width =.3mm,rounded corners, minimum width=\width, minimum height= \height, draw=charcoal] (\nodenum) at (\xpos,\ypos) {\name};
    
    \renewcommand{\xpos}{-.2*\xrel}
	\draw [decorate,line width=.75pt,color=charcoal,decoration={brace,amplitude=5pt},xshift=\xrel,yshift=-\rowspace]	(0*\xrel+\xglobalshift,2.5*\rowspace+\yrel+\yglobalshift) --++ (3.5*\xrel,0) node [black,midway,xshift=9pt] {};
	
    \renewcommand{\nodenum}{v1}
    \renewcommand{\name}{\large Layer $2$}
	\renewcommand{\xpos}{1.75*\xrel+\xglobalshift}
    \renewcommand{\ypos}{3*\rowspace+\yrel+\yglobalshift}
    \renewcommand{\height}{\heightdouble}
    \renewcommand{\width}{\widthsingle}
    \node[] (\nodenum) at (\xpos,\ypos) {\name};

    \renewcommand{\xpos}{-.2*\xrel}
	\draw [decorate,line width=.75pt,color=charcoal,decoration={brace,amplitude=5pt},xshift=\xrel,yshift=-\rowspace]	(-4.5*\xrel+\xglobalshift, 2.5*\rowspace+1*\yrel+\yglobalshift) --++ (3.5*\xrel,0) node [black,midway,xshift=9pt] {};
 
    \renewcommand{\nodenum}{v1}
    \renewcommand{\name}{\large Layer $1$}
	\renewcommand{\xpos}{-2.75*\xrel+\xglobalshift}
    \renewcommand{\ypos}{3*\rowspace+\yrel+\yglobalshift}
    \renewcommand{\height}{\heightdouble}
    \renewcommand{\width}{\widthsingle}
    \node[] (\nodenum) at (\xpos,\ypos) {\name};

    \renewcommand{\nodenum}{v9}
    \renewcommand{\name}{\LARGE $\cdots$}
	\renewcommand{\xpos}{4.5*\xrel+\xglobalshift}
    \renewcommand{\ypos}{1*\rowspace+\yglobalshift}
    \renewcommand{\height}{\heightsingle}
    \renewcommand{\width}{\widthsingle}
    \node[] (\nodenum) at (\xpos,\ypos) {\name};

\end{tikzpicture}
}
\end{center}
\caption{(a) Numerical simulations (blue points) of $\insta_N$ for a unitary $U_k$ as in (c, top), which is a single layer of $k$ T gates. Black points are the exact values of the $\insta$: $\insta(U_k)=k\log(4/3)$. The system size is $n=10$, the Pauli sample complexity is 500 and the OTOC sample complexity is 500. (b) Experimental measurement of $\insta_N$ for $U_k$. The system size is $n=5$, the Pauli sample complexity is 500, and the OTOC sample complexity is 500. (d) Experimental measurement of $\insta_N$ for the unitary $V_k$ in (c, bottom), which contains $k$ layers. Layer $i$ is composed of a layer of H gates, two layers of staggered CNOT gates, a layer of S gates, and a single T gate applied to the $i$-th qubit. The system size is $n=4$, the Pauli sample complexity is $500$ and the OTOC sample complexity is $500$. In all plots, each data point is computed by independently measuring $\insta$ (or $\insta_N$) 5 times and averaging. The OTOC is measured using the circuit in Fig.~\ref{Fig:Circuit}.} 
\label{Fig:Plots}
\end{figure*}
Corollary~\ref{Cor_insta} shows that circuits with small (large) quantities of magic can (cannot) be efficiently measured. To illustrate, we consider the unitary ${U_k=T^{\otimes k}\otimes I^{\otimes n-k}}$. Using Theorem~\ref{Thm:SamplesComplexity}, we find ${N=e^{8k/3}f(\eta,\delta)}$. The sample complexity grows exponentially with the number of T gates, $k$, in the circuit. The measurement is efficient when $k=log(n)$ and is intractable when $k=linear(n)$.

In the case of more complex circuit architectures, one still typically expects the sample complexity to grow exponentially with the number of T gates. This is because magic typically scales with the number of T gates. This suggests that one can only measure the magic of circuits which are classically simulable (i.e., those containing $log(n)$ T gates). These circuits cannot demonstrate a quantum advantage. This motivates the following conjecture.

\begin{conjecture}\label{conjecture1}
One cannot efficiently and accurately measure a magic monotone $\mathcal{M}$ when $\mathcal{M}=linear(n)$.
\end{conjecture}
In other words, we conjecture that measuring $\mathcal{M}$ for a circuit with a $linear(n)$ number of T gates is intractable (assuming that $\mathcal{M}$ scales linearly with the number of T gates). Informally, this conjecture captures the idea that, as more T gates are added to a circuit, measuring any reliable magic monotone should become harder. This is because the sample complexity is expected to scale exponentially with the amount of magic (or alternatively, with the number of T gates). This is relevant to random quantum circuits, which typically contain large quantities of magic~\cite{PRXQuantum.3.020333, PhysRevB.107.035148, Chen_Garcia_Bu_Jaffe_2024}.

We sketch the following argument to support the conjecture. Many magic monotones which scale linearly with the number of $T$ gates (or T states), $N_T$, are defined in terms of logarithms. They can, for example, take on the form ${\mathcal{M}=-\log(exp(-N_T))}$. The magic entropy~\cite{bu2023stabilizer}, stabilizer R\'enyi entropy~\cite{PhysRevLett.128.050402}, and additive Bell magic~\cite{PRXQuantum.4.010301} are some examples. Accurately extracting the $exp(-N_T)$ value requires a measurement error exponentially small in $N_T$, leading to a sample complexity exponentially large in $N_T$. This makes the measurement intractable for large $N_T$ (i.e., when $N_T=linear(n)$), consistent with the conjecture. In the Supplemental Information, we show that the additive Bell magic satisfies Conjecture~\ref{conjecture1}.

Thus far, we have considered the complexity of Pauli sampling. We must also consider the number of physical measurements needed to measure the OTOC, which we call the \textit{OTOC sample complexity}. Many OTOC measurement protocols have been constructed. Past examples have utilized: an interferometric approach~\cite{PhysRevA.94.040302}, a randomized measurement toolbox~\cite{PhysRevX.9.021061, Elben2023}, a teleportation-based technique~\cite{Landsman2019, PhysRevX.11.021010}, and the classical shadows formalism~\cite{PhysRevResearch.3.033155, Huang_2020}. Here, we use the method proposed by Swingle et al.~\cite{PhysRevA.94.040302}, given by the circuit in Fig.~\ref{Fig:Circuit} (we give an alternative circuit in the Supplemental Information which uses only $n+1$ qubits). By running this circuit a finite number of times, we can approximate the OTOC up to a given error. The following proposition gives the OTOC sample complexity.

\begin{prop}[OTOC sample complexity]\label{Prop:OTOC}
With probability at least $1-\delta$, the number of samples needed to accurately measure $\OTOC(U,P_1,P_2)$ up to an error of $\gamma\OTOC(U,P_1,P_2)$ (where $0<\gamma<1$) is
\begin{equation}
    M=\frac{\ln(1/\delta)}{\gamma^2 \OTOC(U,P_1,P_2)^2}.
\end{equation}
\end{prop}

\begin{figure}[h!]
\begin{center}
\scalebox{.46}{
\begin{tikzpicture}
\renewcommand{\xglobalshift}{0}
\renewcommand{\yglobalshift}{-6*\rowspace}
    \draw [thick,color=charcoal](0*\xrel+\xglobalshift,\yrel+2*\rowspace+\yglobalshift)--++(15.5*\xrel, 0)
    (0*\xrel+\xglobalshift,\yrel+\rowspace+\yglobalshift)--++(15.5*\xrel, 0)
    (0*\xrel+\xglobalshift,\yrel+\yglobalshift)--++(15.5*\xrel, 0)
    (0*\xrel+\xglobalshift,\yrel-\rowspace+\yglobalshift)--++(15.5*\xrel, 0)
    (0*\xrel+\xglobalshift,\yrel-2*\rowspace+\yglobalshift)--++(14.5*\xrel, 0);

    \draw [thick,color=charcoal](14.25*\xrel+\xglobalshift,1.25*\yrel-2*\rowspace+\yglobalshift)--++(1*\xrel, 0)
(14.25*\xrel+\xglobalshift,.75*\yrel-2*\rowspace+\yglobalshift)--++(1*\xrel, 0);

    \renewcommand{\nodenum}{v1}
    \renewcommand{\name}{\Large $\ket{0}$}
	\renewcommand{\xpos}{-0.75*\xrel+\xglobalshift}
    \renewcommand{\ypos}{2*\rowspace+\yrel+\yglobalshift}
    \renewcommand{\height}{\heightsingle}
    \renewcommand{\width}{\widthsingle}
    \node[] (\nodenum) at (\xpos,\ypos) {\name};
    
    \renewcommand{\nodenum}{v1}
    \renewcommand{\name}{\Large $\ket{0}$}
	\renewcommand{\xpos}{-0.75*\xrel+\xglobalshift}
    \renewcommand{\ypos}{1*\rowspace+\yrel+\yglobalshift}
    \renewcommand{\height}{\heightsingle}
    \renewcommand{\width}{\widthsingle}
    \node[] (\nodenum) at (\xpos,\ypos) {\name};
    
    \renewcommand{\nodenum}{v1}
    \renewcommand{\name}{\Large $\ket{0}$}
	\renewcommand{\xpos}{-0.75*\xrel+\xglobalshift}
    \renewcommand{\ypos}{0*\rowspace+\yrel+\yglobalshift}
    \renewcommand{\height}{\heightsingle}
    \renewcommand{\width}{\widthsingle}
    \node[] (\nodenum) at (\xpos,\ypos) {\name};
    
    \renewcommand{\nodenum}{v1}
    \renewcommand{\name}{\Large $\ket{0}$}
	\renewcommand{\xpos}{-0.75*\xrel+\xglobalshift}
    \renewcommand{\ypos}{-1*\rowspace+\yrel+\yglobalshift}
    \renewcommand{\height}{\heightsingle}
    \renewcommand{\width}{\widthsingle}
    \node[] (\nodenum) at (\xpos,\ypos) {\name};
    
    \renewcommand{\nodenum}{v1}
    \renewcommand{\name}{\Large $\ket{+}_{\mathcal{C}}$}
	\renewcommand{\xpos}{-0.75*\xrel+\xglobalshift}
    \renewcommand{\ypos}{-2*\rowspace+\yrel+\yglobalshift}
    \renewcommand{\height}{\heightsingle}
    \renewcommand{\width}{\widthsingle}
    \node[] (\nodenum) at (\xpos,\ypos) {\name};

    \renewcommand{\xpos}{-1.5*\xrel+\xglobalshift}
	\draw [decorate,line width=.75pt,color=charcoal,decoration={brace,amplitude=5pt},xshift=\xrel,yshift=-\rowspace]	(\xpos, 1*\rowspace+\yrel+\yglobalshift) -- ++(0, 1*\rowspace) node [black,midway,xshift=9pt] {};
 
    \renewcommand{\xpos}{-1.5*\xrel+\xglobalshift}
	\draw [decorate,line width=.75pt,color=charcoal,decoration={brace,amplitude=5pt},xshift=\xrel,yshift=-\rowspace]	(\xpos, -1*\rowspace+\yrel+\yglobalshift) -- (\xpos,1*\yrel+\yglobalshift) node [black,midway,xshift=9pt] {};

    \renewcommand{\nodenum}{v1}
    \renewcommand{\name}{\Large Ref}
	\renewcommand{\xpos}{-2.25*\xrel+\xglobalshift}
    \renewcommand{\ypos}{1.5*\rowspace+\yrel+\yglobalshift}
    \renewcommand{\height}{1*\heightsingle}
    \renewcommand{\width}{1*\widthsingle}
    \node[] (\nodenum) at (\xpos,\ypos) {\name};

    \renewcommand{\nodenum}{v1}
    \renewcommand{\name}{\Large Sys}
	\renewcommand{\xpos}{-2.25*\xrel+\xglobalshift}
    \renewcommand{\ypos}{0*\rowspace+\yglobalshift}
    \renewcommand{\height}{1*\heightsingle}
    \renewcommand{\width}{1*\widthsingle}
    \node[] (\nodenum) at (\xpos,\ypos) {\name};
    
    \renewcommand{\nodenum}{v1}
    \renewcommand{\name}{\Large $H$}
	\renewcommand{\xpos}{.75*\xrel+\xglobalshift}
    \renewcommand{\ypos}{2*\rowspace+\yrel+\yglobalshift}
    \renewcommand{\height}{1*\heightsingle}
    \renewcommand{\width}{1*\widthsingle}
    \node[rectangle, fill=white,  line width =.3mm,rounded corners, minimum width=\width, minimum height= \height, draw=charcoal] (\nodenum) at (\xpos,\ypos) {\name};
 
    \renewcommand{\nodenum}{v1}
    \renewcommand{\name}{\Large $H$}
	\renewcommand{\xpos}{.75*\xrel+\xglobalshift}
    \renewcommand{\ypos}{1*\rowspace+\yrel+\yglobalshift}
    \renewcommand{\height}{1*\heightsingle}
    \renewcommand{\width}{1*\widthsingle}
    \node[rectangle, fill=white,  line width =.3mm,rounded corners, minimum width=\width, minimum height= \height, draw=charcoal] (\nodenum) at (\xpos,\ypos) {\name};
    
    %CNOT gates
    \renewcommand{\xpos}{1.75*\xrel+\xglobalshift}
    \renewcommand{\ypos}{2*\rowspace+\yrel+\yglobalshift}
    \renewcommand{\height}{.2*\heightsingle}
    \node[circle, fill=charcoal, line width =.3mm, minimum height=\height,  draw=charcoal] (\nodenum) at (\xpos,\ypos) {};

    \draw [thick,color=charcoal]
    (1.75*\xrel+\xglobalshift, 2*\rowspace+\yrel+\yglobalshift)--++(0,-4.3*\yrel);
    
	\renewcommand{\xpos}{1.75*\xrel+\xglobalshift}
    \renewcommand{\ypos}{0*\rowspace+\yrel+\yglobalshift}
    \renewcommand{\height}{.2*\heightsingle}
    \node[circle,  line width =.3mm, minimum height=\height,  draw=charcoal] (\nodenum) at (\xpos,\ypos) {};	
       
    %CNOT gates
    \renewcommand{\xpos}{2.75*\xrel+\xglobalshift}
    \renewcommand{\ypos}{1*\rowspace+\yrel+\yglobalshift}
    \renewcommand{\height}{.2*\heightsingle}
    \node[circle, fill=charcoal, line width =.3mm, minimum height=\height,  draw=charcoal] (\nodenum) at (\xpos,\ypos) {};

    \draw [thick,color=charcoal]
    (2.75*\xrel+\xglobalshift, 1*\rowspace+\yrel+\yglobalshift)--++(0,-4.3*\yrel);
    
	\renewcommand{\xpos}{2.75*\xrel+\xglobalshift}
    \renewcommand{\ypos}{-1*\rowspace+\yrel+\yglobalshift}
    \renewcommand{\height}{.2*\heightsingle}
    \node[circle,  line width =.3mm, minimum height=\height,  draw=charcoal] (\nodenum) at (\xpos,\ypos) {};	

    %CNOT gates
	\renewcommand{\xpos}{4*\xrel+\xglobalshift}
    \renewcommand{\ypos}{-2*\rowspace+\yrel+\yglobalshift}
    \renewcommand{\height}{.2*\heightsingle}
    \node[circle, fill=charcoal,  line width =.3mm, minimum height=\height,  draw=charcoal] (\nodenum) at (\xpos,\ypos) {};

    \draw [thick,color=charcoal]
    (4*\xrel+\xglobalshift, -2*\rowspace+\yrel+\yglobalshift)--++(0,2.3*\yrel);
	
    \renewcommand{\nodenum}{v1}
    \renewcommand{\name}{\Large $P_2$}
	\renewcommand{\xpos}{4*\xrel+\xglobalshift}
    \renewcommand{\ypos}{\yglobalshift}
    \renewcommand{\height}{3*\heightsingle}
    \renewcommand{\width}{1.5*\widthsingle}
    \node[rectangle, fill=white,  line width =.3mm,rounded corners, minimum width=\width, minimum height= \height, draw=charcoal] (\nodenum) at (\xpos,\ypos) {\name};

    \renewcommand{\nodenum}{v1}
    \renewcommand{\name}{\Large $U$}
	\renewcommand{\xpos}{5.5*\xrel+\xglobalshift}
    \renewcommand{\ypos}{\yglobalshift}
    \renewcommand{\height}{3*\heightsingle}
    \renewcommand{\width}{1.5*\widthsingle}
    \node[rectangle, fill=white,  line width =.3mm,rounded corners, minimum width=\width, minimum height= \height, draw=charcoal] (\nodenum) at (\xpos,\ypos) {\name};

    \renewcommand{\nodenum}{v1}
    \renewcommand{\name}{\Large $P_1$}
	\renewcommand{\xpos}{7*\xrel+\xglobalshift}
    \renewcommand{\ypos}{\yglobalshift}
    \renewcommand{\height}{3*\heightsingle}
    \renewcommand{\width}{1.5*\widthsingle}
    \node[rectangle, fill=white,  line width =.3mm,rounded corners, minimum width=\width, minimum height= \height, draw=charcoal] (\nodenum) at (\xpos,\ypos) {\name};

    \renewcommand{\nodenum}{v1}
    \renewcommand{\name}{\Large $U^\dagger $}
	\renewcommand{\xpos}{8.5*\xrel+\xglobalshift}
    \renewcommand{\ypos}{\yglobalshift}
    \renewcommand{\height}{3*\heightsingle}
    \renewcommand{\width}{1.5*\widthsingle}
    \node[rectangle, fill=white,  line width =.3mm,rounded corners, minimum width=\width, minimum height= \height, draw=charcoal] (\nodenum) at (\xpos,\ypos) {\name};

    \renewcommand{\nodenum}{v1}
    \renewcommand{\name}{\Large $X$}
	\renewcommand{\xpos}{9.75*\xrel+\xglobalshift}
    \renewcommand{\ypos}{-2*\rowspace+\yrel+\yglobalshift}
    \renewcommand{\height}{\heightsingle}
    \renewcommand{\width}{\widthsingle}
    \node[rectangle, fill=white,  line width =.3mm,rounded corners, minimum width=\width, minimum height= \height, draw=charcoal] (\nodenum) at (\xpos,\ypos) {\name};

    %CNOT gates
	\renewcommand{\xpos}{11*\xrel+\xglobalshift}
    \renewcommand{\ypos}{-2*\rowspace+\yrel+\yglobalshift}
    \renewcommand{\height}{.2*\heightsingle}
    \node[circle, fill=charcoal,  line width =.3mm, minimum height=\height,  draw=charcoal] (\nodenum) at (\xpos,\ypos) {};

    \draw [thick,color=charcoal]
    (11*\xrel+\xglobalshift, -2*\rowspace+\yrel+\yglobalshift)--++(0,2.3*\yrel);
	
    \renewcommand{\nodenum}{v1}
    \renewcommand{\name}{\Large $P_2$}
	\renewcommand{\xpos}{11*\xrel+\xglobalshift}
    \renewcommand{\ypos}{\yglobalshift}
    \renewcommand{\height}{3*\heightsingle}
    \renewcommand{\width}{1.5*\widthsingle}
    \node[rectangle, fill=white,  line width =.3mm,rounded corners, minimum width=\width, minimum height= \height, draw=charcoal] (\nodenum) at (\xpos,\ypos) {\name};
    
    \renewcommand{\nodenum}{v1}
    \renewcommand{\name}{\Large $X$}
	\renewcommand{\xpos}{12.25*\xrel+\xglobalshift}
    \renewcommand{\ypos}{-2*\rowspace+\yrel+\yglobalshift}
    \renewcommand{\height}{\heightsingle}
    \renewcommand{\width}{\widthsingle}
    \node[rectangle, fill=white,  line width =.3mm,rounded corners, minimum width=\width, minimum height= \height, draw=charcoal] (\nodenum) at (\xpos,\ypos) {\name};

    \renewcommand{\nodenum}{v1}
    \renewcommand{\name}{\Large $H$}
	\renewcommand{\xpos}{13.25*\xrel+\xglobalshift}
    \renewcommand{\ypos}{-2*\rowspace+\yrel+\yglobalshift}
    \renewcommand{\height}{\heightsingle}
    \renewcommand{\width}{\widthsingle}
    \node[rectangle, fill=white,  line width =.3mm,rounded corners, minimum width=\width, minimum height= \height, draw=charcoal] (\nodenum) at (\xpos,\ypos) {\name};
    
    \renewcommand{\nodenum}{v1}
    \renewcommand{\name}{\includegraphics[scale=0.075]{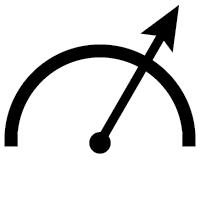}}
	\renewcommand{\xpos}{14.25*\xrel+\xglobalshift}
    \renewcommand{\ypos}{-2*\rowspace+\yrel+\yglobalshift}
    \renewcommand{\height}{\heightsingle}
    \renewcommand{\width}{\widthsingle}
    \node[rectangle, fill=white,  line width =.3mm,rounded corners, minimum width=\width, minimum height= \height, draw=charcoal] (\nodenum) at (\xpos,\ypos) {\name};

    \renewcommand{\nodenum}{v1}
    \renewcommand{\name}{}
	\renewcommand{\xpos}{13.775*\xrel+\xglobalshift}
    \renewcommand{\ypos}{-2*\rowspace+\yrel+\yglobalshift}
    \renewcommand{\height}{1.7*\heightsingle}
    \renewcommand{\width}{3.1*\widthsingle}
    \node[rectangle, dashed, line width =.3mm,rounded corners, minimum width=\width, minimum height= \height, draw=charcoal] (\nodenum) at (\xpos,\ypos) {\name};

\end{tikzpicture}
}
\end{center}
\caption{Quantum circuit to measure the OTOC for a unitary $U$ and Pauli strings $P_1$ and $P_2$. `Ref' denotes $n$ reference qubits and `Sys' denotes $n$ system qubits. The control qubit is in the state $\ket{+}_{\mathcal{C}}=(\ket{0}_{\mathcal{C}}+\ket{1}_{\mathcal{C}})/\sqrt{2}$. The dotted box denotes a measurement in the $X$ basis. The circuit outputs the expectation value $\langle X_\mathcal{C} \rangle=\OTOC(U, P_1, P_2)$, where $\mathcal{C}$ denotes the control qubit.}
\label{Fig:Circuit}
\end{figure}
Proposition~\ref{Prop:OTOC} shows that measuring smaller OTOC values requires more samples. The following corollary gives the regime where an accurate measurement is possible.

\begin{corollary}\label{Cor:OTOC}The OTOC can be efficiently and accurately approximated when $\OTOC(U)=\frac{1}{poly(n)}$. However, an accurate approximation is intractable when $\OTOC(U)=exp(-n)$.
\end{corollary}
In the case of Haar random unitaries, the value of the OTOC is typically $exp(-n)$. This renders the OTOC measurement intractable, as the required sample complexity is exponentially large in $n$. Furthermore, introducing more magic into a quantum circuit can decrease the magnitude of the OTOC to near 0, as shown in~\cite{Mi_2021}. By Proposition~\ref{Prop:OTOC}, a greater number of samples are subsequently required for an accurate measurement, demonstrating the effect of magic on measurement precision.

\section{Simulations and Experiments}
We numerically simulate a measurement of magic to understand how the measurement accuracy changes with the number of $T$ gates in a circuit. In Fig.~\ref{Fig:Plots}\ (a), we simulate $\insta_N$ for the unitary $U_k$ in Fig.~\ref{Fig:Plots}\ (c, top), which contains $k$ T gates. Each data point is computed with $n=10$ qubits and a Pauli sample complexity of $N=500$. This is much smaller than the $16^{10}$ samples required to compute the exact value of Pauli instability using Eq.~\eqref{Eq:Pauli_instability}. 

The simulated data initially scales linearly with the number of T gates, agreeing with the exact value. When the number of $T$ gates is on the same order as the system size (10 in this case), the accuracy of the approximation breaks down. This is notably seen after adding 5 or more T gates. This shows that we require more samples to accurately approximate Pauli instability as more magic is introduced into the circuit. This is consistent with Theorem~\ref{Thm:SamplesComplexity}. 

We experimentally measure magic on the IBM Eagle quantum processor. We perform the experiment on a small scale system of 4 to 5 qubits to mitigate noise effects. In Fig.~\ref{Fig:Plots}\ (b), we measure $\insta_N$ for the unitary $U_k$ on a 5-qubit system (red points). The inherent noise in the processor results in the experimental values initially overestimating the true values (black points) and simulated values (blue points). The experimental values gradually approach the exact values, before giving an underestimate at 5 T gates. The simulated values also underestimate the exact values as the magic increases. These results again showcase that when magic is large, we require more samples for an accurate estimate. 

To explain why noise can lead to an overestimate, we provide a simple example. Assume that $U_k$ is subject to depolarizing noise with a strength of $\lambda$. Pauli instability becomes ${\insta(U_k)\rightarrow \insta(U_k) -\mathrm{log}(1-\lambda)}$. As the noise increases, the monotone increases, giving a false signature of magic, consistent with our data (notably, the first two red data points in Fig.~\ref{Fig:Plots}\ (b)).

In practice, circuits used in quantum computation have more complicated architectures than $U_k$. Namely, they contain multiple layers of Clifford and non-Clifford gates which generate entanglement~\cite{HorodeckiRMP09, ChitambarRMP19, PhysRevLett.70.1895, Islam2015}. We experimentally verify that our monotone can capture an increase in magic as T gates are added to more complex circuits. This property verifies the monotone's reliability. In {Fig.~\ref{Fig:Plots}\ (d)}, we experimentally measure the magic of the circuit architecture, $V_k$, found in Fig.~\ref{Fig:Plots}\ (c, bottom). This circuit is composed of $T$ gates and entangling layers of Clifford gates. The plot displays a roughly linear relation with the number of T gates, similar to Fig.~\ref{Fig:Plots}\ (b). This agrees with the intuition that a monotone should typically increase as more T gates are added. Noise effects become more prevalent as the circuit depth increases, leading to larger experimental values of the monotone, compared to the simulated values.

\section{Conclusion}
We have introduced a scalable measure of magic for large-scale quantum computers. Our measure allows us to prove that small quantities of magic (i.e., in the regime where there is no quantum advantage) can be approximated on large quantum devices. When the magic is extensive, measurement becomes intractable. We conjecture this to be true for any reliable magic measure. We further conjecture that measuring magic when a quantum computer exhibits a quantum advantage is hard. We leave the proof of these statements as an open problem. 

Our result can be interpreted as a precision problem. As more magic is introduced into a quantum computer, a greater measurement precision is required, making our task more difficult. This is reminiscent of the barren plateau problem in quantum machine learning, where ultra-fine measurement precision prevents the training of certain learning models~\cite{McClean2018, Garcia2023}. As a consequence, only models providing no quantum advantage have been shown to be trainable~\cite{Cong2019, Cerezo2021, cerezo2024does}. Our results lead us to pose the following problem: can one prove that magic cannot generally be learned via quantum machine learning? We postulate that one should encounter a barren plateau. 

We have shown, both theoretically and experimentally, that noise can lead to a false signature of magic. As quantum processors are inherently noisy, it is useful to develop measurement protocols which are robust to noise. Previous works have successfully measured chaos in the presence of noise~\cite{Landsman2019}. We expect that one can use similar techniques to measure magic robustly. 

%The LOTOC can be interpreted as a measure of operator entanglement, which quantifies chaos. It has been claimed that operator entanglement is hard to simulate classically~\cite{Mi_2021}. A natural question is: can one efficiently measure a resource monotone which quantifies operator entanglement in chaotic systems? We suspect not, as operator entanglement is generated by magic. Nevertheless, it may also be interesting to explore whether the LOTOC, or other magic monotones, can be used to bound the simulation cost of operator entanglement.

\section{Acknowledgements}
We thank Dolev Bluvstein for discussions on magic. We thank Sieglinde Pfaendler for discussions on qiskit. This work was supported in part by the ARO Grant W911NF-19-1-0302, the ARO
MURI Grant W911NF-20-1-0082, and the NIH grant GM136859.

\bibliography{Bibliography}
%BIBLIOGRAPHY DOESN'T WORK IF THERE ARE SPACES IN MAIN DOCUMENT TITLE 

\end{document}